\documentclass[12pt]{article}

\usepackage{amsfonts,amssymb, amsmath,mathrsfs}

\usepackage[english]{babel}

\def\on#1#2{\mathop{\vbox{\ialign{##\crcr\noalign{\kern2pt}
$\scriptstyle{#2}$\crcr\noalign{\kern2pt\nointerlineskip}
\kern-2pt$\hfil\displaystyle{#1}\hfil$\crcr}}}\limits}

\def\e{{\rm e}}

\def\nn{ \nonumber }
\def\bq{ \begin{equation} }
\def\eq{ \end{equation} }
\def\ben{ \begin{eqnarray} }
\def\en{ \end{eqnarray} }

\newtheorem{opr}{Definition}

\newtheorem{prop}{Proposition}

\newtheorem{exa}{Example}
\newenvironment{exam}{\begin{exa} \rm }{\end{exa}}

\newtheorem{remark}{Remark}
\newenvironment{rem}{\begin{remark} \rm}{\end{remark}}

\begin{document}


\title{On the two different bi-Hamiltonian structures for the Toda lattice.}

\author{A V Tsiganov \\
\it\small
St.Petersburg State University, St.Petersburg, Russia\\
\it\small e--mail: tsiganov@mph.phys.spbu.ru}
\date{}
 \maketitle

\begin{abstract}
We construct two different incompatible Poisson
pencils for the Toda lattice by using known variables of
separation proposed by Moser and by Sklyanin.
\end{abstract}

\section{Introduction}
\setcounter{equation}{0} A bi-Hamiltonian manifold $M$ is a smooth manifold endowed with two compatible bi-vectors $P_0,P_1$ such that
\[
[\![P_0,P_0]\!]=[\![P_0,P_1]\!]=[\![P_1,P_1]\!]=0,\] where $[\![.,.]\!]$ is the Schouten bracket.
Such a condition assures that the linear combination $P_0-\lambda P_1$ is a Poisson pencil, i.e. it is a Poisson bi-vector for each $\lambda \in \mathbb C$, and therefore the corresponding bracket $\{.,.\}_0-\lambda\{.,.\}_1$ is a pencil of Poisson brackets \cite{mag97}.

Dynamical systems on $M$ having enough functionally independent integrals of motion $H_1,\ldots H_{n}$ in involution with respect to the both Poisson brackets
\bq\label{bi-ham}
\{H_i,H_j\}_0=\{H_i,H_j\}_1=0\,.
\eq
will be called bi-integrable systems.

The main aim of this note is to prove that any separable system is bi-integrable system. As an example we show that the Toda lattice is bi-integrable system with respect to two different Poisson pencils $P_0+\lambda P_1$ and $P_0+\lambda P^{\,\star}_1$, which are related with two known families of the separated variables \cite{mos75,skl85a}.

\section{The separation of variables method}
\setcounter{equation}{0}

A complete integral $S(q,t)$ of the
Hamilton-Jacobi equation
\bq\label{Eq-HJt} \dfrac{\partial S(q,t)}{\partial
t}+H\left(q,\dfrac{\partial S(q,t)}{\partial q},t\right)=0\,,
\eq
where $q=(q_1,\ldots,q_n)$, is a solution $S(q,t,\alpha_1,\ldots,\alpha_n)$ depending on $n$ parameters $\alpha=(\alpha_1,\ldots, \alpha_n)$
such that
\bq\label{m-cond}
\det\left\|\dfrac{\partial ^2 S(q,t,\alpha)}{\partial q_i\partial
\alpha_j}\right\|\neq 0\,.
\eq
For any complete integral of (\ref{Eq-HJt}) solutions
$q_i=q_i(t,\alpha,\beta)$ and $p_i=p_i(t,\alpha,\beta)$ of the
Hamilton equations of motion may be found from the Jacobi equations
\bq \label{Eq-J}
\beta_i=-\dfrac{\partial S(q,t,\alpha)}{\partial
\alpha_i},\qquad p_i=\dfrac{\partial S(q,t,\alpha)}{\partial q_i}\,,\qquad
i=1,\ldots,n.
\eq
According to (\ref{m-cond}) if we resolve a second part of the Jacobi equations with respect to parameters $\alpha_1,\ldots, \alpha_n$ one gets $n$ independent integrals of motion
\bq\label{m-int}
\alpha_m=H_m(p,q,t),\qquad m=1,\ldots,n,
\eq
as functions on the phase space $M$ with coordinates $p,q$.

 \begin{opr}
A dynamical system is a separable system if the corresponding complete integral $S(q,t,\alpha)$
has an additive form
 \bq\label{Add-Int}
S(q,t,\alpha_1,\ldots,\alpha_n)= -Ht+\sum_{i=1}^n
S_i(q_i,\alpha_1,\ldots,\alpha_n).
\eq
Here the $i$-th component $S_i$ depends only on the $i$-th
coordinate $q_i$ and $\alpha$.
\end{opr}
In such a case Hamiltonian $H$ is said to be separable and coordinates $q$ are said to be separated coordinates for $H$, in order to stress that the possibility to find an additive complete integral of (\ref{Add-Int}) depends on the choice of the coordinates.

For the separable dynamical system we have
\bq
\dfrac{\partial^2
}{\partial q_k\partial q_j}S=\dfrac{\partial
}{\partial q_k}\,\left(\dfrac{\partial
}{\partial q_j}S_j\right)=0,\qquad\mbox{\rm for all}\qquad j\neq k,
\eq
such that the second Jacobi equations (\ref{Eq-J}) are the separated equations
\bq\label{Eq-Js} p_j=\dfrac{\partial
}{\partial q_j}S_j(q_j,\alpha_1,\ldots,\alpha_n)\,,\quad\mbox{\rm or}\quad
\phi_j(p_j,q_j,\alpha)=p_j-\dfrac{\partial
}{\partial q_j}S_j(q_j,\alpha)=0\,.
\eq
Here $j$-th equation contains a pair of canonical variables $p_j$ and $q_j$ only.

\begin{prop}
For any separable dynamical system integrals of motion $H_m(p,q,t)$ (\ref{m-int}) are in the involution
 \[
 \{H_k,H_m\}_f=0,\qquad k,m=1,\ldots,n,
 \]
 with respect to the following brackets $\{.,.\}_f$ on $M$
\bq
\label{poi-f}
\{p_i,q_j\}_f=\delta_{ij}\,f_j(p,q)\,,\qquad\{p_i,p_j\}_f=\{q_i,q_j\}_f=0,
\eq
which depend on $n$ arbitrary functions $f_1(p,q),\ldots,f_n(p,q)$.
\end{prop}
\textsc{Proof:} In fact we have to repeat the proof of the Jacobi theorem due by Liouville.
Namely, differentiate the relations (\ref{m-int}) by $q_j$ and then substitute momenta $p_k$ from separated equations (\ref {Eq-Js}) to obtain
\[
\dfrac{\partial H_m
}{\partial q_j}+\sum_{i=1}^n \dfrac{\partial H_m
}{\partial p_i}\dfrac{\partial p_i}{\partial q_j}=\left(
\dfrac{\partial H_m
}{\partial q_j}+\dfrac{\partial H_m
}{\partial p_j}\dfrac{\partial^2 S_j
}{\partial q_j^2}\right)=0\,.
\]
It follows that for any $H_k$ and $H_m$
\[
\sum_{j=1}^n f_j\dfrac{\partial H_k
}{\partial p_j}\left(
\dfrac{\partial H_m
}{\partial q_j}+\dfrac{\partial H_m
}{\partial p_j}\dfrac{\partial^2 S_j
}{\partial q_j^2}
\right)=
\sum_{j=1}^n f_j\dfrac{\partial H_k
}{\partial p_j}
\dfrac{\partial H_m
}{\partial q_j}
+\sum_{j=1}^n f_j\dfrac{\partial H_k
}{\partial p_j}\dfrac{\partial H_m
}{\partial p_j}\dfrac{\partial^2 S_j
}{\partial q_j^2}=0\,.
\]
Permute indexes $k$ and $m$ and subtract the resulting equation from the previous one we get
\[
\sum_{j=1}^n f_j\left(\dfrac{\partial H_k
}{\partial p_j}
\dfrac{\partial H_m
}{\partial q_j}
-\dfrac{\partial H_m
}{\partial p_j}
\dfrac{\partial H_k
}{\partial q_j}\right)=0\,.
\]
The final assertion easily follows.
\vskip0.3cm
Brackets $\{.,.\}_f$ (\ref{poi-f}) are the Poisson brackets if and only if
\bq\label{poi-f2}
[\![P_f,P_f]\!]=0,
\eq
where
\bq
\label{f-ten}
P_f=\left(
 \begin{array}{cc}
 0 & \mbox{\rm diag}(f_1,\ldots,f_n) \\
 -\mbox{\rm diag}(f_1,\ldots,f_n) & 0
 \end{array}
 \right)\,.
\eq
\begin{prop}
If $j$-th function
\bq\label{f-eq}
f_j(p,q)=f_j(p_j,q_j),\qquad j=1,\ldots,n,
\eq
depends only on the $j$-th pair of coordinates $p_j,q_j$ then brackets $\{.,.\}_f$ (\ref{poi-f}) are the Poisson brackets, which are compatible with canonical ones.
\end{prop}
\textsc{Proof:} Substitute tensor $P_f$ (\ref{f-ten}) into the equations
\[
[\![P_0,P_f]\!]=[\![P_f,P_f]\!]=0,\qquad{\mbox{ \rm where}}\qquad
P_0=\left(
 \begin{array}{cc}
 0 & \mathrm I \\
 -\mathrm I & 0
 \end{array}
 \right),
\]
one gets the following system of partial differential equations
\[
\dfrac{\partial f_j}{\partial q_k}=\dfrac{\partial f_j}{\partial p_k}=0,\qquad
f_i\dfrac{\partial f_j}{\partial q_k}=f_i\dfrac{\partial f_j}{\partial p_k}
=0,\]
for all $j\neq k\neq i$. The separable functions $f_j(p_j,q_j)$ (\ref{f-eq}) satisfy this system of equations. So, the corresponding tensor $P_f$ (\ref{f-ten})
 is the Poisson tensor, which is compatible with canonical tensor $P_0$.
\vskip0.3cm
\begin{rem}
Separated variables $(p,q)$ are defined up to canonical transformations
\[
p_j=X(p_j,q_j),\qquad q_j=Y(p_j,q_j),
\]
which have to preserve canonical tensor $P_0$ and would change the second tensor $P_f$.
It is clear, that freedom in the choice of the functions
$f_j(p_j,q_j)$ is related with this freedom in the definition
of the separated variables.
\end{rem}

According to the following proposition,
the separation of variables method is closely related with the bi-Hamiltonian geometry.
\begin{prop}
Any separable dynamical system is bi-integrable system.
\end{prop}
\textsc{Proof:} In order to get
a pair of compatible Poisson brackets on the phase space $M$ it is enough to postulate
brackets (\ref{poi-f})
\bq\label{poi-f3}
\{p_i,q_j\}_f=\delta_{ij}\,f_j(p_j,q_j),\qquad\{p_i,p_j\}_f=\{q_i,q_j\}_f=0,
\eq
between the known separated variables $q_j$ and $p_j$ for a given dynamical system.
According to the Proposition 1 the corresponding integrals of motion
$H_m(p,q,t)$ (\ref{m-int}) are in the bi-involution with respect to brackets $\{.,.\}_0$ and $\{.,.\}_f$.
\vskip0.3cm
In the next sections we demonstrate how this proposition works
for the open and periodic Toda lattice.
\begin{rem}
The separated variables $(p,q)$ are canonical variables, which put the
corresponding recursion operator $N=P_fP_0^{-1}$ in diagonal form. So, according to \cite{fmp01,fp02},
a set of the separated variables $(p,q)$ is said to be Darboux-Nijenhuis coordinates.
\end{rem}

\begin{rem}
For the stationary systems differentiate the separated equations (\ref{Eq-Js})
\bq\label{f-m}
\left(\dfrac{\partial\phi_j}{\partial q_j}dq_j+
\dfrac{\partial\phi_j}{\partial p_j}dp_j\right)+
\sum_{i=1}^n\dfrac{\partial\phi_j}{\partial H_i}dH_i=0,
\eq
then apply $N^*=P_0^{-1}P_f$ and substitute (\ref{f-m})
in the resulting equation one gets
\ben
f_j\left(\dfrac{\partial\phi_j}{\partial q_j}dq_j+
\dfrac{\partial\phi_j}{\partial p_j}dp_j\right)&+&
\sum_{i=1}^n\dfrac{\partial\phi_j}{\partial H_i}N^*dH_i=\nn\\
&=&
-f_j\sum_{i=1}^n\dfrac{\partial\phi_j}{\partial H_i}dH_i+
\sum_{i=1}^n\dfrac{\partial\phi_j}{\partial H_i}N^*dH_i
=0\,.\nn
\en
It follows that
\[
\sum_{i=1}^n\dfrac{\partial\phi_j}{\partial H_i}N^*dH_i=f_j\sum_{i=1}^n\dfrac{\partial\phi_j}{\partial H_i}dH_i,\qquad j=1,\ldots, n,
\]
that is, in matrix form,
\bq\label{F-pedr}
N^*dH=dH\,F.
\eq
Here $n\times n$ control matrix $F$ with eigenvalues $f_1,\ldots,f_n$
is defined by
\[
F=J^{-1}\mbox{\rm diag}(f_1,\ldots,f_n)\,J,\qquad J_{ji}=\dfrac{\partial\phi_j}{\partial H_i}
\]
and $dH$ is $2n\times n$ matrix with entries $
dH_{ij}={\partial H_j}/{\partial z_i}$, where $z=(q,p)$.

Equation (\ref{F-pedr}) means that the subspace spanned by covectors $dH_1,\ldots,dH_n$ is invariant with respect to $N^*$
\cite{fp02}.
\end{rem}

\begin{exam}

For further use we introduce
the special control matrix $F$, which is the Fr\"obenius matrix
\bq\label{f-num}
 F_f=\left(%
 \begin{matrix}
 c_1 & 1 & 0 &\cdots & 0 \\
 c_2 & 0 & 1 & \cdots &0 \\
 \vdots & & \cdots &0 & 1 \\
 c_n &0 & \cdots & &0
 \end{matrix}%
 \right).
 \eq
 Here $c_k$ are coefficients of the characteristic polynomial
 $\Delta_N(\lambda)$ of the recursion operator $N=P_fP_0^{-1}$:
 \bq\label{delt-N}
\Delta_N(\lambda)=\Bigl(\det(N-\lambda\, {\mathrm
I})\Bigr)^{1/2}=\lambda^n-(c_1\lambda^{n-1}+\ldots+c_n)=\prod_{j=1}^n
(\lambda-\lambda_j)\,.
\eq
\end{exam}

\section{Open Toda lattice}
Let us consider open Toda associated with the root system of
$\mathscr A_n$ type. The Hamilton function is equal to
\[
H=\dfrac12\sum_{i=1}^n {p_i}^2+\sum_{i=1}^{n-1} \e^{q_i-q_{i+1}},
\]
where $q_i$ denotes the position of $i$-th particle and $p_i$ is its momenta, such that
\bq\label{darb} \{q_i,p_j\}_0=\delta_{ij},\qquad
\{p_i,p_j\}_0=\{q_i,q_j\}_0=0\,.
\eq
Consequently, the equations of motion read
\[
\dot{q}_i=p_i,\qquad \dot{p}_1=-\e^{q_1-q_2},\qquad \dot{p}_n=\e^{q_{n-1}-q_{n}}
\]
and
\[\dot{p}_i=\e^{q_{i-1}-q_i}-\e^{q_i-q_{i+1}},\qquad i=2,\ldots,n-1\,.\]
The exact solution is due to existence of a Lax matrix. Consider the $L$-operator
\bq L_i=\left(\begin{array}{cc}
 \lambda-p_i &\, -e^{q_i} \\
 e^{-q_i}& 0
\end{array}\right)\,,
\eq
and the monodromy matrix
\bq\label{22toda}
T(\lambda)=L_1(\lambda)\cdots L_{n-1}(\lambda)\,L_n(\lambda)=
\left(\begin{array}{cc}
A(\lambda)& B(\lambda) \\
C(\lambda)& D(\lambda)
\end{array}\right),\qquad \det T(\lambda)=1,
\eq
which depends polynomially on
the parameter $\lambda$
\bq
T(\lambda)=\left(\begin{array}{ll}
\lambda^n+A_1\lambda^{n-1}+\ldots +A_n\qquad& B_1\lambda^{n-1}+\ldots+B_n \\
C_1\lambda^{n-1}+\ldots+C_n& D_2\lambda^{n-2}+\ldots+D_n
\end{array}\right).\nn
\eq
The monodromy matrix satisfies Sklyanin's Poisson brackets:
\bq
\{{T}(\lambda)\on{,}{\otimes}{T}(\mu)\}_0= [r(\lambda-\mu),\,
{T}(\lambda)\otimes {T}(\mu)\,]\,, \label{rrpoi}
\eq
where
$r(\lambda-\mu)$ is the $4\times4$ rational $r$-matrix
\bq
r(\lambda-\mu)=\dfrac{-1}{\lambda-\mu}\Pi,\qquad \Pi=\left(\begin{array}{cccc}
 1 & 0 & 0 & 0 \\
 0 & 0 & 1 & 0 \\
 0 & 1 & 0 & 0 \\
 0 & 0 & 0 & 1
\end{array}\right)\,,\quad \eta\in {\mathbb C}\,.\label{rr}
\eq

Monodromy matrix $T(\lambda)$ is the Lax matrix for
periodic Toda lattice, whereas the Lax matrix for
open Toda lattice is equal to
\[
T_{o}(\lambda)=KT(\lambda)=\left(\begin{array}{cc}
A& B \\
0& 0
\end{array}\right)(\lambda),\qquad K=\left(\begin{array}{cc}
1& 0 \\
0& 0
\end{array}\right)\,.
\]
The trace of the Lax matrix
\bq\label{int-open}
\mbox{tr}\,T_o(\lambda)=A(\lambda) = \lambda^n +H_1\lambda^{n-1}+\cdots H_n,\qquad \{H_i,H_j\}=0\,.
\eq
generates $n$ independent integrals of motion $H_i$ in the involution providing complete integrability of the system \cite{skl85a}.

\subsection{The Moser variables}
According to \cite{mos75} we introduce the
$n$ pairs of separated variables $\lambda_i$, $\mu_i$, $i=1,\ldots,n$, having the standard Poisson brackets,
\bq
\left\{\lambda_i,\lambda_j\right\}_0=\left\{\mu_i,\mu_j\right\}_0=0,\qquad \left\{\lambda_i,\mu_j\right\}_0=\delta_{ij},
\label{Darb}
\eq
with the $\lambda$-variables being $n$ zeros of the polynomial $A(\lambda)$ and the $\mu$-variables being values of the polynomial $B(\lambda)$ at those zeros,
\bq
A(\lambda_i)=0,\qquad \mu_i=\eta^{-1}\ln B(\lambda_i),\qquad i=1,\ldots,n.
\label{dn-var}
\eq
The Moser variables $\lambda_j$ may be identified with poles of
the Baker-Akhiezer function $\vec\Psi$ associated with
the Lax matrix $T(\lambda)$ (\ref{22toda})
\[T(\lambda)\vec\Psi=\lambda\vec\Psi,\qquad (\vec\Psi,\vec{\alpha})=const,\]
having some very special normalization $\vec{\alpha}$ \cite{skl95}.

The interpolation data (\ref{dn-var}) plus $n$ identities
\[B(\lambda_i)C(\lambda_i)=\det T(\lambda)=1\]
allow us to construct the separation representation for the whole monodromy matrix $T(\lambda)$:
\bq
\label{DN-rep}
\begin{array}{l}
A(\lambda)=(\lambda-\lambda_1)(\lambda-\lambda_2)\cdots(\lambda-\lambda_n),\\
\\
B(\lambda)=A(\lambda)\sum_{i=1}^n \dfrac{\e^{\mu_i}}{(\lambda-\lambda_i)A^{'}(\lambda_i)},\\
\\
C(\lambda)=-A(\lambda)\sum_{i=1}^n \dfrac{\e^{-\mu_i}}
{(\lambda-\lambda_i)A^{'}(\lambda_i)},\\
\\
D(\lambda)=\dfrac{1+B(\lambda)C(\lambda)}{A(\lambda)}\,.
\end{array}
\eq
If we postulate the following second Poisson brackets (\ref{poi-f3})
\[
\{\lambda_i,\mu_j\}_1=\lambda_i\delta_{ij}, \qquad \{\lambda_i,\lambda_j\}_1=\{\mu_i,\mu_j\}_1=0
\]
one gets \cite{ts06a}
\bq\label{br-AC}
\begin{array}{l}
\{A(\lambda),A(\mu)\}_1=\{B(\lambda),B(\mu)\}_1=0\,,\\
\\
\{A(\lambda),B(\mu)\}_1=\dfrac{1}{\lambda-\mu}\Bigl(\mu\, A(\lambda)B(\mu)-\lambda\, A(\mu)B(\lambda)\Bigr)\,.
\end{array}
\eq
The first bracket in (\ref{br-AC}) guaranties that integrals of motion $H_i$ (\ref{int-open}) from $A(\lambda)$ are in the bi-involution.

Substitute polynomials $A(\lambda)$ and $B(\lambda)$ in initial $(p,q)$-variables into the brackets (\ref{br-AC}) and solve the resulting equations to obtain the known second Poisson tensor \cite{das89,fern93}
\bq\label{toda-gen} P_1^{open}=\sum_{i=1}^{n-1}
e^{q_i-q_{i+1}}\dfrac{\partial}{\partial
p_{i+1}}\wedge\dfrac{\partial}{\partial p_{i}} +\sum_{i=1}^n
p_i\dfrac{\partial}{\partial q_{i}}\wedge\dfrac{\partial}{\partial
p_{i}}+\sum_{i<j}^n \dfrac{\partial}{\partial
q_{j}}\wedge\dfrac{\partial}{\partial q_{i}}.
\eq
The minimal characteristic polynomial of the
corresponding recursion operator $N_M=P_1^{open}P_0^{-1}$ is equal to
\[\Delta_{N_M}(\lambda)=A(\lambda)=\prod_{j=1}^n(\lambda-\lambda_j)=\lambda^n+\sum_{j=1}^n H_j\lambda^{n-j}\,.\]
So, $(\lambda,\mu)$-coordinates are variables of separation of the action-angle type \cite{fp02}, i.e. the corresponding separated equations are trivial
\[\{H_i,\lambda_j\}=0\,,\qquad i,j=1,\ldots,n.\]

The Hamiltonians $H_i$ (\ref{int-open}) from $A(\lambda)$
satisfy the Fr\"obenius recursion relations
\bq\label{fr-ch1}
N^*_M\,dH_i=dH_{i+1}-H_i\,dH_1,
\eq
where $N^*_M=P_0^{-1}P_1^{open}$ and $H_{n+1}=0$, i.e. Hamiltonians $H_i$ satisfy equation (\ref{F-pedr}) with the Fr\"obenius matrix $F_f$ (\ref{f-num}).

\begin{exam}
For the 3-particles open Toda lattice
Hamiltonians $H_i$ (\ref{int-open}) are
\ben
H_1&=&-(p_1+p_2+p_3),\nn\\
H_2&=&p_1p_2+p_1p_3+p_2p_3-\e^{q_1-q_2}-\e^{q_2-q_3},
\nn\\
H_3&=&-p_1p_2p_3+p_1\e^{q_2-q_3}+p_3\e^{q_1-q_2}\,.\nn
\en
It's obvious that $H=H_1^2/2-H_2$. The second Poisson tensor $P_1$ (\ref{toda-gen}) in the matrix form reads as
\bq\label{pp3}
P_1^{open}=\left(\begin{array}{cccccc}
0& -1& -1& p_1& 0& 0\\
1& 0& -1& 0& p_2& 0\\
1& 1& 0& 0& 0& p_3\\
-p_1& 0& 0& 0& -\e^{q_1-q_2}& 0\\
0& -p_2& 0& \e^{q_1-q_2}& 0& -\e^{q_2-q_3}\\
0& 0& -p_3& 0& \e^{q_2-q_3}& 0
\end{array}\right).
\eq
The control matrix $F$ in (\ref{F-pedr}) is the Fr\"obenius matrix
\bq\label{frob-mat}
F_M^{open}=\left(
 \begin{array}{ccc}
 -H_1 & 1 & 0 \\
 -H_2 & 0 & 1 \\
 -H_3 & 0 & 0
 \end{array}
 \right),
\eq
where coefficients $c_i=-H_i$ coincide with integrals of motion.
\end{exam}

\subsection{The Sklyanin variables}
 According to \cite{skl85a, skl95} we can consider another set of
the separated coordinates, which are poles of the Baker-Akhiezer function $\vec\Psi$
associated with the Lax matrix $T(\lambda)$ (\ref{22toda}) having the
standard normalization $\vec\alpha=(0,1)$.

In this case the first half of
variables is coming from $(n-1)$ finite roots and logarithm of
leading coefficient of the non-diagonal entry of the monodromy
matrix
\bq\label{B-int}
B(\lambda)=-e^{u_n}\prod_{j=1}^{n-1}(\lambda-v_j)\,,
\eq
Another half is given by
 \bq \label{toda-per} u_j=-\ln
A(v_j),\quad j=1,\ldots,n-1,\qquad\mbox{\rm and}\qquad
v_n=\sum_{i=1}^n p_i\,.
\eq
In these separated variables other entries of $T(\lambda)$ read as
\bq\label{A-int}
A(\lambda)=\left(\lambda+\sum_{j=1}^{n}v_j\right)\prod_{j=1}^{n-1}(\lambda-v_j)+
\sum_{j=1}^{n-1} e^{-u_j}\prod_{i\neq
j}^{n-1}\dfrac{\lambda-v_i}{v_j-v_i}\,.
\eq
and
\[
D(\lambda)=-\sum_{j=1}^{n-1} e^{u_j}\prod_{i\neq
j}^{n-1}\dfrac{\lambda-v_i}{v_j-v_i}\,,\qquad C(\lambda)=\dfrac{A(\lambda)D(\lambda)-1}{B(\lambda)}.
\]

If we postulate the second Poisson brackets (\ref{poi-f3})
\[
\{v_i,u_j\}_1^\star=v_i\delta_{ij}, \qquad \{v_i,v_j\}_1^\star=\{u_i,u_j\}_1^\star=0
\]
one gets
\bq\label{AA2}
\{A(\lambda),A(\mu)\}_1^\star=\{B(\lambda),B(\mu)\}_1^\star=0\,,\\
\eq
and
\ben
\{A(\lambda),B(\mu)\}_1^\star&=&\dfrac{1}{\lambda-\mu}\Bigl(\lambda\, A(\lambda)B(\mu)-\mu\, A(\mu)B(\lambda)\Bigr)\nn\\
&+&{\e^{-u_n}}\left(\lambda+\mu+\sum_{i=1}^{n-1}v_i\right)B(\lambda)B(\mu)\,.
\en
In initial $(p,q)$-variables the last bracket looks like
\ben
\{A(\lambda),B(\mu)\}_1^\star&=&\dfrac{1}{\lambda-\mu}\Bigl(\lambda\, A(\lambda)B(\mu)-\mu\, A(\mu)B(\lambda)\Bigr)\label{AB2}\\
&+&{\e^{-q_n}}\left(\lambda+\mu+\sum_{i=1}^{n-1}p_i\right)B(\lambda)B(\mu)\,.\nn
\en
The first bracket in (\ref{AA2}) guaranties that integrals of motion $H_i$ (\ref{int-open}) from $A(\lambda)$ are in the bi-involution.

Substitute into the brackets (\ref{AA2})-(\ref{AB2}) polynomials $A(\lambda)$ and $B(\lambda)$ in initial $(p,q)$-variables and solve the resulting equations to obtain the following Poisson tensor
\ben\label{toda-gen2} P^{\,\star}_1&=&\sum_{i=1}^{n-2}
e^{q_i-q_{i+1}}\dfrac{\partial}{\partial
p_{i+1}}\wedge\dfrac{\partial}{\partial p_{i}} +\sum_{i=1}^{n-1}
p_i\dfrac{\partial}{\partial q_{i}}\wedge\dfrac{\partial}{\partial
p_{i}}+\sum_{i<j}^{n-1} \dfrac{\partial}{\partial
q_{j}}\wedge\dfrac{\partial}{\partial q_{i}}\\
&+&\sum_{i=1}^{n-1}(p_i+\mathrm p)\dfrac{\partial}{\partial p_n}\wedge \dfrac{\partial}{\partial q_i}+\sum_{i=2}^{n-1}\left(\e^{q_i-q_{i+1}}-\e^{q_{i-1}-q_i}\right)\dfrac{\partial}{\partial p_n}\wedge \dfrac{\partial}{\partial p_i}\nn\\
&+&\mathrm p\dfrac{\partial}{\partial p_n}\wedge \dfrac{\partial}{\partial q_n}+
\e^{q_1-q_2}\dfrac{\partial}{\partial p_1}\wedge \dfrac{\partial}{\partial p_n}
+
\e^{q_{n-2}-q_{n-1}}\dfrac{\partial}{\partial p_{n-1} }\wedge \dfrac{\partial}{\partial p_n}\,,\nn
\en
where $\mathrm{p}=\sum_{i=1}^n p_i$ is a total momentum.

The tensor $P^{\,\star}_1$ is independent on $q_n$ and
the minimal characteristic polynomial of the corresponding recursion operator $ N_S=P^{\,\star}_1 P_0^{-1}$ is equal to
\[
\Delta_{N_S}=-e^{q_n}(\lambda+\mathrm p)B(\lambda)\,.
\]
The normalized traces of the powers of $N_S$ are integrals of motion for $n-1$ particle open Toda lattice. As consequence the Hamiltonians $H_i$ (\ref{int-open}) from $A(\lambda)$
satisfy equation (\ref{F-pedr}) with the following control matrix
\bq\label{f-open2}
F_S^{open}=\left(
 \begin{array}{cc}
 -\mathrm p & 0 \\
 0 & F_M^{open} \\
 \end{array}
 \right),
\eq
where $F_M^{open}$ is the Fr\"obenius matrix (\ref{f-num}) associated with the recursion operator $N_M$
for $n-1$ particle open Toda lattice.

The corresponding separated equations follow directly from
the definitions of $(u,v)$-variables (\ref{toda-per})
\bq\label{tod-seq}
e^{-u_j}-A(v_j)=0\,,\quad j=1,\ldots,n-1\qquad\mbox{\rm and}\qquad
\Phi_n=v_n-\alpha_1=0\,,
\eq
where $A(\lambda)=\lambda^n+\sum_{i=1}^{n-1}\alpha_i\lambda^{n-i}$ and $\alpha_i=H_i$ are the values of integrals of motion.
The first $(n-1)$ separated equations give rise to the equations of motion
\bq\label{toda-eqm}
\{A(\lambda),v_j\}=A(v_j)\,\prod_{i\neq
j}^{n-1}\dfrac{\lambda-v_i}{v_j-v_i}\,,\qquad j=1,\ldots,n-1,
\eq
which are linearized by the Abel transformation \cite{skl85a}
\[\left\{A(\lambda),\sum_{k=1}^{n-1} \int^{v_k} \sigma_{j}\right\}=-\lambda^{j-1},\qquad
\sigma_j=\frac{\lambda^{j-1}\,d\lambda }{A(\lambda)},\qquad
j=1,\ldots,n-1\,,
\]
where $\{\sigma_j\}$ is a basis of abelian differentials of first
order on an algebraic curve $z=A(\lambda)$ corresponding to the
separated equations (\ref{tod-seq}).

\begin{rem}
From the factorization of the monodromy
matrix $T(\lambda)$ (\ref{22toda}) one gets
\[
 T_n(\lambda)=T_{n-1}(\lambda)L_n(\lambda),\qquad \Rightarrow\qquad
 B_{n}(\lambda)=-\e^{q_{n}}A_{n-1}(\lambda),
\]
where $B_n(\lambda)$ is entry of the monodromy matrix $T_n(\lambda)$ of $n$ particles Toda lattice
and $A_{n-1}(\lambda)$ is entry of the monodromy matrix $T_{n-1}(\lambda)$ of $n-1$ particles Toda lattice.

This implies that for the $(n-1)$-particle chain the Moser variables $\lambda_j$ coincide with the $n-1$ Sklyanin variables $u_j$, $i=1,\ldots,n-1$ for the $n$-particle chain.
\end{rem}

\begin{exam}
At $n=3$ the Poisson tensor $P^{\,\star}_1$ in the matrix form reads as
\[
P^{\,\star}_1=\left(\begin{array}{cccccc}
0& -1& 0& p_1& 0& -p_1-p\\
1& 0& 0& 0& p_2& -p_2-p\\
0& 0& 0& 0& 0& p\\
-p_1& 0& 0& 0& -\e^{q_1-q_2}& \e^{q_1-q_2}\\
0& -p_2& 0& \e^{q_1-q_2}& 0& -\e^{q_1-q_2}\\
p_1+p& p_2+p& p& -e^{q_1-q_2}& \e^{q_1-q_2}& 0
\end{array}\right).
\]
and at n=4 it looks like
\[
P^{\,\star}_1=\left(\begin{smallmatrix}
0& -1& -1& 0& p_1& 0& 0& -p_1-p\\
1& 0& -1& 0& 0& p_2& 0& -p_2-p\\
1& 1& 0& 0& 0& 0& p_3 &-p_3-p\\
0& 0& 0& 0& 0& 0& 0& p\\
-p_1& 0& 0& 0& 0& -\e^{q_1-q_2}& 0& \e^{q_1-q_2}\\
0& -p_2& 0& 0& \e^{q_1-q_2}& 0& -\e^{q_2-q_3}& \e^{q_2-q_3}-\e^{q_1-q_2}\\
0& 0& -p_3& 0& 0& \e^{q_2-q_3}& 0&\e^{q_2-q_3}\\
p_1+p&p_2+p& p_3+p&p&-\e^{q_1-q_2}& -\e^{q_2-q_3}+\e^{q_1-q_2}&-\e^{q_2-q_3}&0
\end{smallmatrix}\right).
\]
The corresponding control matrices $F$ in (\ref{F-pedr}) are given by
\bq\label{f-33op}
F_S^{open}=\left(
 \begin{array}{ccc}
 -p_1-p_2-p_3 & 0 & 0 \\
 0 & p_1+p_2 & 1 \\
 0 & -p_1p_2+\e^{q_1-q_2} & 0
 \end{array}
 \right).
\eq
and
\bq
F_S^{open}=\left(
 \begin{smallmatrix}
 -p_1-p_2-p_3-p_4 & 0 & 0 & 0 \\
 \\
 0 & p_1+p_2+p_3 & 1 & 0 \\
 \\
 0 & -p_1p_2-p_1p_3-p_2p_3+\e^{q_1-q_2}-\e^{q_2-q_3} & 0 & 1 \\
 \\
 0 & p_1p_2p_3-p_1\e^{q_2-q_3}-p_3\e^{q_1-q_2} & 0 & 0
 \end{smallmatrix}
 \right)\,.
\eq

\end{exam}

\subsection{Periodic Toda lattice}
The trace of the Lax matrix $T(\lambda)$ (\ref{22toda}) for periodic Toda lattice
\bq\label{int-cl}
\mbox{tr}\,T(\lambda)=A(\lambda)+D(\lambda) = \lambda^n +H_1\lambda^{n-1}+\cdots H_n,\qquad \{H_i,H_j\}=0\,.
\eq
generates $n$ independent integrals of motion $H_i$ in the involution providing complete integrability of the periodic Toda lattice \cite{skl85a}. For instance the Hamilton function is equal to
\[
H=\dfrac12\sum_{i=1}^n {p_i}^2+\sum_{i=1}^{n-1} \e^{q_i-q_{i+1}}+\e^{q_n-q_1}.
\]

The Moser $(\lambda,\mu)$-variables (\ref{dn-var}) are the separated variables for the open Toda lattice only. Nevertheless, the bi-Hamiltonian structure for the periodic Toda lattice may be constructed by using generic properties of the Sklyanin bracket (\ref{rrpoi}) \cite{ts07a}. According to \cite{ts07a}, in order to get the second Poisson tensor $P^{per}_1$ for the periodic Toda lattice
we have to apply canonical transformation
\bq\label{toda-trans}
p_1\to p_1+\e^{-q_1},\qquad p_n\to p_n+\e^{q_n}\,.
\eq
to the tensor $P_1^{open}$ (\ref{toda-gen}) for the open Toda lattice .
In this case Hamiltonians (\ref{int-cl}) satisfy equation (\ref{F-pedr}) with the standard Fr\"obenius matrix $F_f$ (\ref{f-num}) associated with the recursion operator $N_M^{per}=P_1^{per}P_0^{-1}$ \cite{ts07a}.

The Sklyanin $(u,v)$-variables (\ref{B-int})-(\ref{toda-per}) are the separated variables for the open and periodic Toda lattice simultaneously. Therefore, the one Poisson tensor $P^{\,\star}_1$ determines the bi-Hamiltonian structure for the both Toda lattices, but the corresponding Hamiltonians satisfy equation (\ref{F-pedr}) with the slightly different control matrices $F^{open}_S$ and $F_S^{per}$.

Namely, for the open Toda lattice the control matrix $F^{open}_S$ is given by (\ref{f-open2}). For the periodic Toda lattice we have to change the first column of this control matrix $F^{open}_S$ (\ref{f-open2}) by the rule:
\bq \label{f-per2}
F_{1,1}^{per}=-\mathrm p,\qquad F_{i,1}^{per}=(\mathrm p+\sum_{j=1}^{n-1}p_j)F_{i,2}^{open}
+F^{open}_{i+1,2},\qquad i=2,\ldots,n\,,
\eq
where $F_{n+1,2}^{open}=0$.

\begin{rem}
For the open and periodic Toda lattices we have two matrix equations
for the same integrals of motion:
\[
N_M^*dH=dH\,F_M,\qquad N_S^*dH=dH\,F_S
\]
So, we have various complimentary relations
\[
N^*_MN^*_SdH=dH\,F_M\,F_S,\qquad{\mbox{\rm or}}\qquad N^*_S(N^*_M)^{-1}dHF_M=dH\,F_S\,.
\]
between integrals of motion $H_i$, recursion operators $N_S$, $N_M$ and the corresponding control matrices $F_M$, $F_S$.

\end{rem}

\begin{exam}For the 3-particles periodic Toda lattice
Hamiltonians $H_i$ (\ref{int-cl}) are
\ben
H_1&=&-(p_1+p_2+p_3),\nn\\
H_2&=&p_1p_2+p_1p_3+p_2p_3-\e^{q_1-q_2}-\e^{q_2-q_3}-\e^{q_3-q_1},
\nn\\
H_3&=&-p_1p_2p_3+p_1\e^{q_2-q_3}+p_2\e^{q_3-q_1}+p_3\e^{q_1-q_2}\,.\nn
\en
The second Poisson tensor (\ref{pp3}) after canonical transformation (\ref{toda-trans}) reads as
\[
{P}_1^{per}=\left(\begin{array}{cccccc}
0& -1& -1& \e^{-q_1}+p_1& 0& \e^{q_3}\\
1& 0& -1& \e^{-q_1}& p_2& \e^{q_3}\\
1& 1& 0& \e^{-q_1}& 0& p_3+\e^{q_3}\\
-\e^{-q_1}-p_1& -\e^{-q_1}& -\e^{-q_1}& 0& -\e^{q_1-q_2}& \e^{q_3-q_1}\\
 0& -p_2& 0& \e^{q_1-q_2}& 0& -\e^{q_2-q_3}\\
 -\e^{q_3}& -\e^{q_3}& -p_3-\e^{q_3}& -\e^{q_3-q_1}& \e^{q_2-q_3}& 0
\end{array}\right)\,.
\]
The corresponding control matrix $F$ in (\ref{F-pedr}) is the Fr\"obenius matrix
\bq\label{frob-mat}
F_M^{per}=\left(
 \begin{array}{ccc}
 c_1 & 1 & 0 \\
 c_2 & 0 & 1 \\
 c_3 & 0 & 0
 \end{array}
 \right),
\eq
where
\[
 \begin{array}{l}
 c_1=p_1+\e^{-q_1}+p_2+p_3+\e^{q_3},\\
 \\
 c_2=-p_1p_2-p_1p_3-p_2p_3-(p_2+p_3)\e^{-q_1}-(p_1+p_2)\e^{q_3}
 +\e^{q_1-q_2}+\e^{q_2-q_3}-\e^{q_1-q_3},\\
 \\
 c_3=((p_3+\e^{q_3})p_2-\e^{q_2-q_3})p_1
+(p_3+\e^{q_3})p_2\e^{-q_1}-p_3\e^{q_1-q_2}-\e^{-q_1+q_2-q_3}-\e^{q_1-q_2+q_3}\,.
 \end{array}
\]
are coefficients of the minimal characteristic polynomial
of the corresponding recursion operator $N_M^{per}=P_1^{per}P_0^{-1}$.

The same integrals of motion $H_i$ for periodic Toda lattice
satisfy equation (\ref{F-pedr}) with another recursion operator $N_S^*= P_0^{-1}P^{\,\star}_1$
and with the following control matrix (\ref{f-open2})-(\ref{f-per2})
\bq\label{f-33per}
F_S^{per}=\left(
 \begin{array}{ccc}
 -p_1-p_2-p_3 & 0 & 0 \\
 \\
 \e^{q_1-q_2}-p_1p_2+(p_1+p_2)(2p_1+2p_2+p_3) & p_1+p_2 & 1 \\
 \\
 (\e^{q_1-q_2}-p_1p_2)(2p_1+2p_2+p_3) & \e^{q_1-q_2}-p_1p_2 & 0
 \end{array}
 \right).
\eq
\end{exam}

\section{Conclusion.}
For the Toda lattice associated with the root system of
$\mathscr A_n$ type we present two different bi-Hamiltonian structures on $M\simeq \mathbb R^{2n}$.
The introduced Poisson tensors $P_1$ (\ref{toda-gen}) and $P^{\,\star}_1$ (\ref{toda-gen2}) are incompatible
\[
[\![P_1,P^{\,\star}_1]\!]\neq 0\,,
\]
while the corresponding recursion operators $N_M=P_1P_0^{-1}$ and $N_S=P^{\,\star}_1 P_0^{-1}$
take the diagonal form in the Moser and the Sklyanin separated variables respectively.

Associated with the tensor $P_1$ brackets (\ref{br-AC}) were rewritten in the $r$-matrix form in \cite{ts07a}. It will be interesting to get the similar $r$-matrix formulation for the brackets (\ref{AA2})-(\ref{AB2}) associated with the tensor $P^{\,\star}_1$.

The research was partially supported by
the RFBR grant 06-01-00140 and grant NSc5403.2006.1.

\end{document}